\documentclass[twocolumn,times,tighten]{aastex631}

\usepackage{graphicx}
\usepackage{amsmath}
\usepackage{amssymb}
\let\tablenum\undefined
\usepackage{siunitx}

\begin{document}

\title{Slow Cosmic-Ray Diffusion in Supersonic and Super-Alfv\'enic Turbulence}

\author[0000-0002-8455-0805]{Yue Hu}
\altaffiliation{NASA Hubble Fellow}
\affiliation{Institute for Advanced Study, 1 Einstein Drive, Princeton, NJ 08540, USA}
\affiliation{Cahill Center for Astronomy and Astrophysics, California Institute of Technology, Pasadena, CA, USA}
\correspondingauthor{Yue Hu}
\email{yuehu@ias.edu}

\begin{abstract}
Extended TeV--PeV gamma-ray halo observations imply cosmic-ray (CR) diffusion that is both substantially slower than the Galactic mean and only weakly anisotropic despite the presence of a large-scale Galactic magnetic field background. We investigate whether such transport can arise in partially ionized source environments, where ion-neutral damping removes the small-scale fluctuations responsible for gyroresonant scattering. Using two-fluid magnetohydrodynamic turbulence simulations with relativistic test-particle tracking, we find that damping increases the parallel diffusion coefficient by approximately two orders of magnitude in trans-sonic turbulence, thereby strongly enhancing field-aligned escape. In supersonic turbulence, shock-associated magnetic fluctuations survive the damping and sustain non-resonant pitch-angle scattering, limiting the increase in parallel diffusion to a factor of a few to ten. At high Alfv\'enic Mach numbers, $M_A\gtrsim5$, the magnetic-field direction decorrelates over the Alfv\'en scale, driving parallel diffusion coefficient $D_\parallel$ and perpendicular diffusion coefficient $D_\perp$ toward equality. For $M_s\simeq10$, $M_A\simeq5$--$10$, and representative parameters for hundred-TeV gamma-ray source environments, we obtain comparable $D_\parallel$ and $D_\perp$, within the observationally inferred range of $\sim10^{27}$--$10^{28}\,\mathrm{cm^2\,s^{-1}}$. These results show that strongly supersonic, highly super-Alfv\'enic turbulence in source environments can sustain slow, nearly isotropic CR transport even when the ion-neutral damping effect is important.
\end{abstract}


\section{Introduction}\label{sec:intro}

Wide-field gamma-ray observatories have revealed extended TeV--PeV emission around pulsars, supernova-remnant environments, and young stellar clusters
\citep{2014ApJ...796..108A,2021ApJ...914..106A,2021Sci...373..425L,2024PhRvL.132m1002C,2025arXiv251105015C,2025ApJ...978L..20S}.
These systems include leptonic pulsar halos, produced by electrons and
positrons escaping from pulsar wind nebulae \citep{2022FrASS...922100F,2022NatAs...6..199L}, and extended hadronic emission around candidate TeVatrons and PeVatrons \citep{2024SciBu..69..449L}, where energetic nuclei interact with the surrounding gas. Despite their different emission mechanisms, both classes provide constraints on local CR transport. First, the diffusion coefficient near $\sim100\,\mathrm{TeV}$ is typically $D\sim10^{27} - 10^{28}\,\mathrm{cm^2\,s^{-1}}$, two to three orders of magnitude below the Galactic value extrapolated to the same energy \citep{2017Sci...358..911A,2021Natur.594...33C,2022PhRvD.105j3007F,2024SciBu..69..449L}. Second, several halos are approximately spherical despite the tens-to-hundreds-pc coherence of the interstellar magnetic field \citep{2012ARA&A..50...29C,2019NatAs...3..776H,2020A&A...641A..11P,2020ApJ...888...96H,2021ApJ...915...67H,2022MNRAS.510.4952L,2023MNRAS.524.2379H,2026ApJ..1005..195T}, indicating that transport is not strongly organized by a single mean-field direction
\citep{2018A&A...612A...1H,2019PhRvD.100l3015D,2023ApJ...944L..29A,2025SCPMA..6879503L}.

These requirements are difficult to satisfy simultaneously. Cosmic ray self-generated Alfv\'en waves can reduce diffusion near CR sources \citep{1971ApJ...170..265S,2018PhRvD..98f3017E,2021ApJ...914L..13S}, but their growth becomes inefficient at TeV--PeV energies because the CR number density is small and ion-neutral damping is strong in cold gas \citep{2004MNRAS.353..550B,2016ApJ...826..166X,2022PhRvD.105l3008M}. Conversely, sub- or trans-Alfv\'enic background turbulence generally yields $D_\perp\ll D_\parallel$ \citep{2008ApJ...673..942Y,2013ApJ...779..140X,2022MNRAS.512.2111H,2022FrP....10.2799L}, so nearly spherical halos require either a favorable magnetic geometry \citep{2019PhRvL.123v1103L} or an additional mechanism that enhances the field stochasticity.

Furthermore, the interstellar medium is only partially ionized \citep{1992pavi.book.....S,Draine_McKee1993,Ferriere2001, 2008ApJ...677.1151L,2010ApJ...720.1612M}. In such environments, neutral--ion decoupling damps the MHD cascade and can substantially modify the magnetic-field structure below the decoupling scale \citep{2015ApJ...810...44X,2016ApJ...826..166X, 2024MNRAS.527.3945H}. The relevance of this damping to CR transport is set by the ordering between the damping scale and the CR gyroradius. Leptonic halos propagating through warm, highly ionized gas, with ionization fractions of order $\xi_i\sim10^{-1}$ \citep{2011piim.book.....D}, are expected to be only weakly affected because the damping scale lies well below the gyroradii of the particles of interest. Extended hadronic emission around TeVatrons and PeVatrons, however, is frequently associated with molecular material and cold neutral gas \citep{2009MNRAS.396.1629G,2015SSRv..188..187S,2021Natur.594...33C}, where $\xi$ can fall $\sim10^{-3}$--$10^{-7}$ \citep{1998ApJ...499..234C,2009A&A...501..619P,2011piim.book.....D}. In such environments, ion-neutral collisions damp Alfv\'enic fluctuations below $\ell_{\rm dec}\sim10^{-2}$--$10^{-1}\,\mathrm{pc}$
\citep{2010MNRAS.406.1201T,2015ApJ...805..118B,2015ApJ...810...44X,2024MNRAS.527.3945H,2025ApJ...994..142H}. These scales overlap the gyroradii relevant to TeV--PeV particles, removing a substantial fraction of the gyroresonant fluctuations \citep{1966ApJ...146..480J,2002ApJ...578L.117Q,2002PhRvL..89B1102Y,2008ApJ...673..942Y,2022FrP....10.2799L} and favoring rapid field-aligned transport. {Despite the importance, the effect of ion-neutral damping on the CR transport is not clear yet.

In this Letter, we use two-fluid (ionized fluid plus neutral fluid) simulations with relativistic test particles to address this gap and the questions of slow diffusion observed in extended TeV--PeV gamma-ray halos. Particularly, we show that supersonic shocks generate strong magnetic fluctuations that minimize the ion-neutral damping. Super-Alfv\'enic motions make the field dynamically subdominant above the Alfv\'en scale, thereby randomizing its direction and isotropizing transport on larger scales. We quantify these effects separately and compare the resulting diffusion coefficients with those inferred around TeV--PeV accelerators.

\begin{figure*}[t]
\centering
\includegraphics[width=0.925\linewidth]{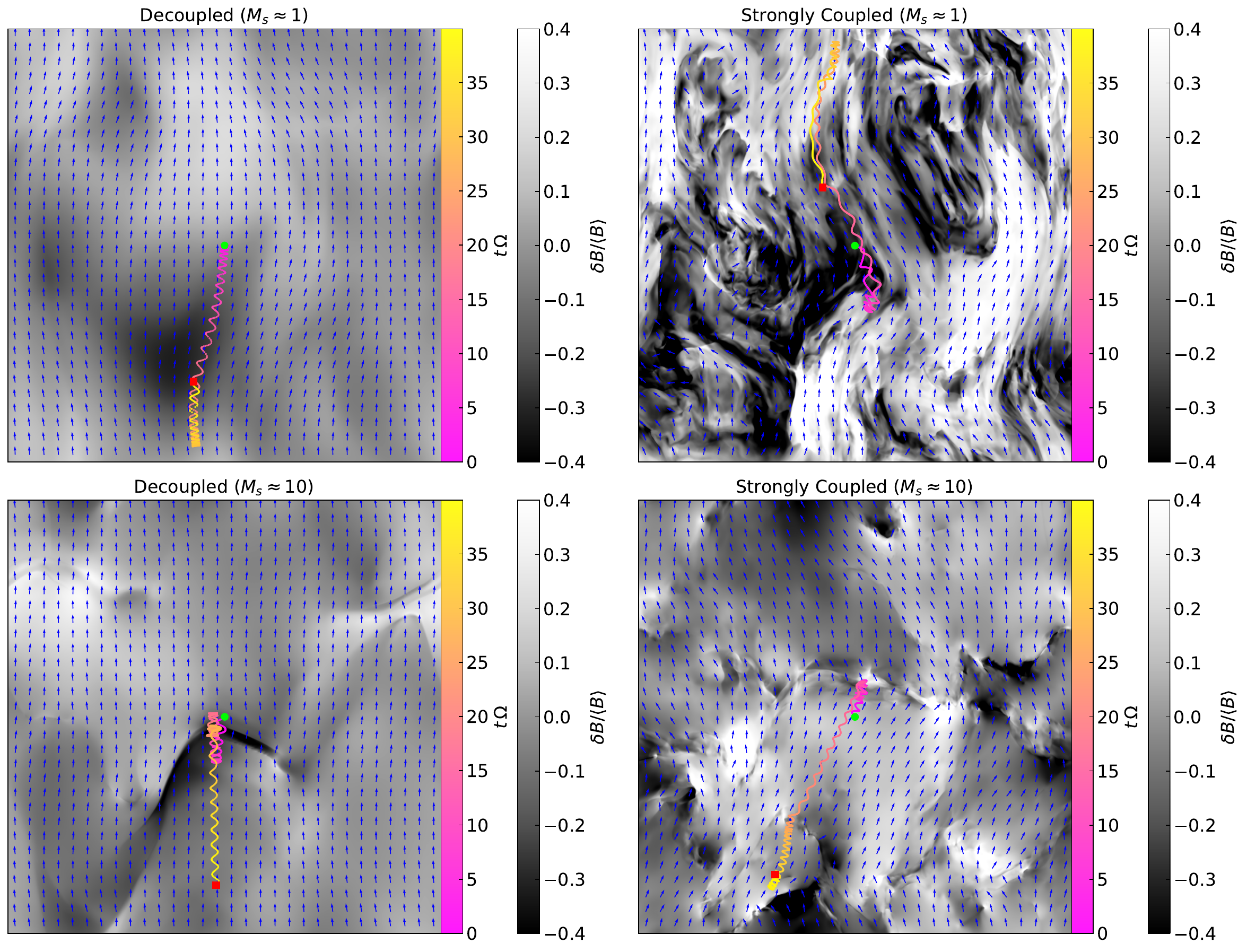}
\caption{\label{fig:map}
Representative CR trajectories with $r_L/L_{\rm box}=8\times10^{-3}$ ($r_L=10$ cells) overlaid on the normalized magnetic fluctuation $\delta B/\langle B\rangle$ in a slice through each particle's initial position. Blue arrows indicate the in-plane magnetic-field direction. The left and right columns show decoupled and strongly coupled two-fluid turbulence, respectively; the upper and lower rows show $M_s\simeq1$ and $M_s\simeq10$, with $M_A\simeq1$ in all panels. Trajectory color denotes $t\Omega$, and green and red markers indicate the initial and final positions.}
\end{figure*}

\section{Numerical Framework}\label{sec:methods}

\subsection{Two-fluid turbulence}
We solve the isothermal two-fluid MHD equations for a charged component, consisting of ions and electrons, and a neutral component in a triply
periodic domain resolved by $1200^3$ cells using \texttt{AthenaK} \citep{2024MNRAS.527.3945H,2026ApJS..283...27S}. The numerica implementation follows \citet{2025ApJ...994..142H}. The domain is threaded by an initially uniform magnetic field, $\boldsymbol{B}_0=B_0\hat{\boldsymbol{z}}$. Turbulence is driven solenoidally at wavenumber $k=2$, corresponding to an injection scale of $L_{\rm inj} = L_{\rm box}/2$.

We characterize the turbulence by the sonic and Alfv\'enic Mach numbers,
\begin{equation}
M_s=\frac{u_{\rm inj}}{c_s},
\qquad
M_A=\frac{u_{\rm inj}}{v_{A}},
\end{equation}
where $u_{\rm inj}$ is the velocity dispersion at the injection scale, $c_s$ is the isothermal sound speed, and $v_{A}$ is the Alfv\'en speed. Our two-fluid runs include trans-sonic ($M_s\simeq1$) and supersonic ($M_s\simeq10$) turbulence, both with $M_A\simeq1$. We also analyze single-fluid ideal-MHD simulations with $M_A=3$, $5$, and $10$ for both $M_s\simeq1$ and $M_s\simeq10$. For the supersonic runs with $M_A\simeq5$ and $10$, we further compare purely solenoidal and purely compressive forcing.

The difference between single-fluid ideal MHD equations and two-fluid equations is the additional momentum exchange term in the momentum equation, represented by the drag force \citep{1992pavi.book.....S}
\begin{equation}
\boldsymbol{f}_{\rm drag}
=\gamma_{\rm d}\rho_i\rho_n
\left(\boldsymbol{u}_n-\boldsymbol{u}_i\right),
\label{eq:drag}
\end{equation}
where $\gamma_{\rm d}$ is the drag coefficient. $\rho$ and $\boldsymbol{u}$ are gas density and velocity, respectively, while the subscripts "$i$" and "$n$" denote the ionized and neutral component.

Neutral--ion decoupling occurs when the neutral--ion collision frequency $\nu_{ni}=\gamma_{\rm d}\rho_i$ falls below the characteristic dynamical frequency of the fluctuations. For Alfv\'en waves, this condition may be written as $\nu_{ni}\lesssim v_A/\ell_{\rm dec}$; in MHD turbulence, the more relevant comparison is with the eddy turnover rate at the corresponding scale \citep{2015ApJ...810...44X}. Here $v_A$ is the Alfv\'en speed and $\ell_{\rm dec}$ denotes the decoupling scale. Below this scale, neutrals no longer follow the ionized component, while the residual ion--neutral drag damps magnetic and velocity fluctuations in the ionized fluid \citep{2015ApJ...810...44X,2024MNRAS.527.3945H}. The damping scale is typically slightly smaller than, but close to, the decoupling scale \citep{2015ApJ...810...44X}; we therefore do not distinguish between the two scales in the following discussion. 

By varying the drag coefficient $\gamma_{\rm d}$, we control the neutral--ion collision frequency and hence the neutral--ion decoupling scale. For both the trans-sonic and supersonic runs, we consider $\frac{\nu_{ni}}{u_{\rm inj}/L_{\rm inj}} = 5\times10^3,\ 5\times10^1,\ \mathrm{and}\ 5,$ which define the strongly coupled, intermediate, and decoupled regimes, respectively. The corresponding decoupling wavenumbers are $k_{\rm dec}\simeq10^4$, $10^2$, and $10$ \citep{2025ApJ...994..142H}. In the strongly coupled case, $k_{\rm dec}\simeq10^4$ lies far beyond the numerical dissipation wavenumber, $k_{\rm dis}\simeq120$, so ions and neutrals remain effectively coupled throughout the resolved inertial and dissipation ranges, recovering the single-fluid limit.


\subsection{Test particles and transport diagnostics}

After the turbulence reaches a statistically stationary state at
$t\gtrsim10L_{\rm inj}/u_{\rm inj}$, we inject $10^3$ relativistic test particles \footnote{ A convergence test is given in the Appendix's Fig.~\ref{fig:convergence}. The two ensembles with $10^3$ and $10^4$ particles exhibit consistent growth and saturation behavior in terms of pitch-angle mean-square variation.} per gyroradius bin with random positions and an initially isotropic pitch-angle distribution. Particle trajectories are integrated in frozen snapshots using a fourth-order Runge--Kutta scheme and the relativistic Lorentz equation. The test-particle approximation is appropriate for the $\sim100\,\mathrm{TeV}$ population considered here because its pressure and number density are insufficient to modify the turbulent fields appreciably. 

The spatial diffusion coefficients are measured from the late-time mean-squared displacements,
\begin{equation}
D_\parallel=\frac{1}{2}\frac{d}{dt}\left\langle\Delta z^2\right\rangle,
\qquad
D_\perp=\frac{1}{4}\frac{d}{dt}
\left\langle\Delta x^2+\Delta y^2\right\rangle,
\label{eq:spatial_diffusion}
\end{equation}
where $z$ is parallel to $\boldsymbol{B}_0$. The pitch-angle diffusion coefficient is $D_{\mu\mu}=(1/2)d\langle(\Delta\mu)^2\rangle/dt$ for particles initialized at $\mu_0$, with $\mu=\boldsymbol{v}\cdot\boldsymbol{B}/(vB)$.

Figs.~\ref{fig:map}--\ref{fig:ma_scan} are presented in dimensionless units based on $cL_{\rm box}$ and the gyrofrequency $\Omega$; only Figs.~\ref{fig:physical_scaling} adopts a dimensional normalization. Because the particle speed $v$, greatly exceeds the turbulent velocity, the particle crossing and scattering timescales are much shorter than the eddy turnover time. We therefore calculate the transport coefficients using a single statistically steady snapshot from each run.

\begin{figure*}[t]
\centering
\includegraphics[width=0.75\linewidth]{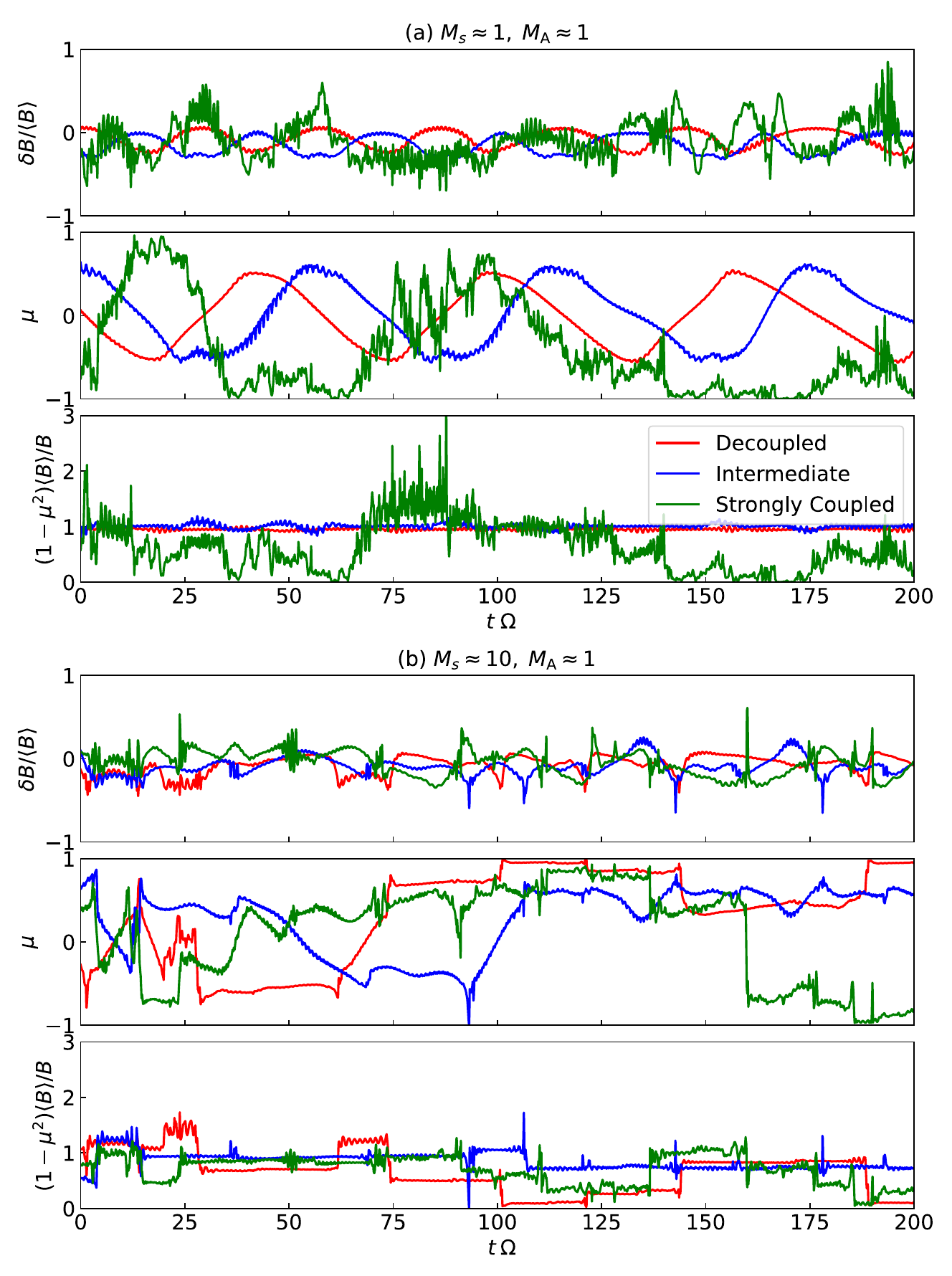}
\caption{\label{fig:trajectory} Evolution along the representative trajectories associated with Fig.~\ref {fig:map}. For $M_s\simeq1$ (left) and $M_s\simeq10$ (right), the panels show $\delta B/\langle B\rangle$, the pitch-angle cosine $\mu$, and the normalized first adiabatic invariant $(1-\mu^2)\langle B\rangle/B$ as functions of $t\Omega$. Red, blue, and green denote decoupled, intermediate, and strongly coupled runs. Approximate conservation of $(1-\mu^2)\langle B\rangle/B$ indicates adiabatic mirror motion, whereas abrupt changes indicate non-adiabatic pitch-angle scattering.}
\end{figure*}

\begin{figure*}[t]
\centering
\includegraphics[width=0.99\linewidth]{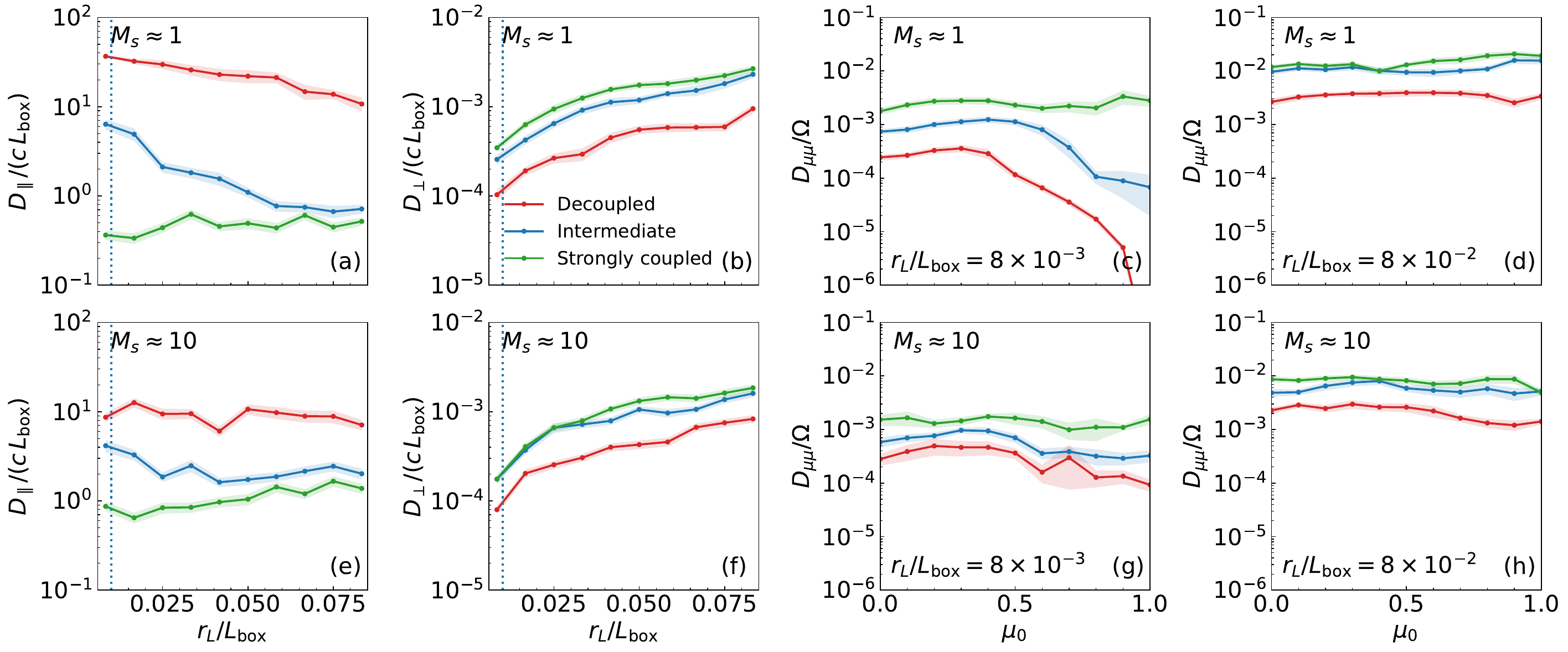}
\caption{\label{fig:coupling}
Dimensionless CR transport coefficients in two-fluid turbulence at $M_A\simeq1$. The upper and lower rows show $M_s\simeq1$ and $M_s\simeq10$, respectively. Panels (a, e) and (b, f) give $D_\parallel/(cL_{\rm box})$ and $D_\perp/(cL_{\rm box})$ as functions of $r_L/L_{\rm box}$. Panels (c, d, g, h) give $D_{\mu\mu}/\Omega$ as a function of $\mu_0$ for $r_L/L_{\rm box}\simeq8\times10^{-3}$ and $8\times10^{-2}$ ($r_L=10$ and $100$ cells). Red, blue, and green denote decoupled, intermediate, and strongly coupled runs; shaded regions indicate statistical uncertainties. The blue dotted lines in panels (a, b, e, f) denote the decoupling scale of the intermediate decoupled run.}
\end{figure*}

\section{Results}\label{sec:results}

\subsection{Magnetic structure under ion-neutral damping}
Fig.~\ref{fig:map} shows how ion-neutral damping modifies the magnetic geomotry sampled by CRs. In the trans-sonic strongly coupled run (upper right), magnetic structure extends throughout the resolved cascade, and the representative trajectory exhibits repeated deflections as the particle encounters fluctuations near and above its gyroradius. In the decoupled run (upper left), power below $\ell_{\rm dec}\simeq L_{\rm box}/10$ is strongly reduced. The plotted particle has $r_L/L_{\rm box}=8\times10^{-3}$, so its resonant scale lies within the damped interval. The remaining field is dominated by comparatively smooth, large-scale fluctuations. Consequently, the trajectory is organized primarily by magnetic mirrors rather than by gyroresonant encounters \citep{2021ApJ...923...53L,2023ApJ...959L...8Z,2025ApJ...994..142H}.

The supersonic cases (lower row) display a different geometry. Both coupling regimes contain narrow magnetic ridges with $\delta B/\langle B\rangle\gtrsim0.4$. Damping reduces the fluctuation power between these structures, but the large-amplitude fluctuations produced by shocks remain prominent. The decoupled supersonic trajectory therefore continues to experience substantial scatterings at these structures. ion-neutral damping remains dynamically relevant, but it does not eliminate the shock-associated magnetic fluctuations generated by supersonic turbulence.

\subsection{Adiabatic mirroring and shock-associated scattering}
Fig.~\ref{fig:trajectory} identifies the corresponding particle dynamics. In the trans-sonic strongly coupled case, $\mu$ reverses repeatedly during the $200$-gyroperiod interval and samples nearly the full range $[-1,1]$. These reversals are accompanied by large departures of $(1-\mu^2)\langle B\rangle/B$ from unity, indicating non-adiabatic pitch-angle scattering with a characteristic time $\tau_s\Omega\sim10$. The intermediate run shows less frequent stochastic changes, consistent with partial removal of the resonant fluctuations.

The decoupled trans-sonic trajectory is qualitatively different. The sampled magnetic fluctuation is smooth, $\mu$ undergoes slow, nearly periodic oscillations with a bounce period $t\Omega\sim50$--$70$, and the normalized adiabatic invariant remains within a few percent of unity. After the small-scale fluctuations are damped, the particle dynamics are therefore dominated by adiabatic reflection between the surviving large-scale mirrors \citep{2021ApJ...923...53L,2023ApJ...959L...8Z,2025ApJ...994..142H}. For a magnetic enhancement from $B$ to $B_{\rm max}$, particles with $|\mu|<\mu_c=[1-B/B_{\rm max}]^{1/2}$are mirror reflected, whereas particles in the loss cone,
$|\mu|>\mu_c$, can escape unless another fluctuation changes their pitch angle. The reduced $D_{\mu\mu}$ in the decoupled run inhibits exchange between these populations: mirror-confined particles undergo slow mirror diffusion, while loss-cone particles propagate rapidly along the field.

For $M_s\simeq10$, damping no longer produces mirror-only dynamics, and shocks provide an additional efficient particle interaction channel \citep{1992MNRAS.255..269B}. The decoupled trajectory encounters magnetic spikes with $\delta B/\langle B\rangle\sim0.3$--$0.5$, and these encounters coincide with abrupt changes in both $\mu$ and the adiabatic invariant. The inferred scattering time, $\tau_s\Omega\sim10$--$20$, is comparable to that of the strongly coupled run. Shock-associated fluctuations therefore maintain impulsive, non-adiabatic scattering without restoring the full small-scale cascade. These encounters continually exchange particles between trapped and passing orbits, preventing the loss-cone population from remaining nearly ballistic for extended periods.

\subsection{Ensemble transport}
Fig.~\ref{fig:coupling} connects the single-particle behavior to ensemble diffusion over the gyroradii from $r_L/L_{\rm box}\simeq8\times10^{-3}$ to $8\times10^{-2}$, corresponding to $r_L=10-100$ cells. In the trans-sonic runs, decoupling increases $D_\parallel/(cL_{\rm box})$ from $0.3$--$0.7$ to $10$--$40$, or by approximately two orders of magnitude. Simultaneously, $D_\perp/(cL_{\rm box})$ decreases from $3\times10^{-4}$--$2\times10^{-3}$ to $10^{-4}$--$7\times10^{-4}$. The resulting anisotropy increases from $D_\parallel/D_\perp\sim10^2$--$10^3$ in the coupled limit to $10^5$--$10^6$ after decoupling. Ion-neutral damping therefore aggravates, rather than alleviates, the conflict with nearly spherical halos in trans-sonic and trans-Alfv\'enic ($M_A\simeq 1$) turbulence.

The pitch-angle coefficients provide the direct explanation. In the strongly coupled trans-sonic run,
$D_{\mu\mu}/\Omega\sim10^{-3}$--$10^{-2}$ over most $\mu_0$. In the decoupled run, $D_{\mu\mu}$ is suppressed by one to several orders of magnitude and falls $\sim10^{-6}$ near $\mu_0\rightarrow1$ for the smaller gyroradius. This is the loss-cone population that escapes most efficiently after the gyroresonant fluctuations are removed. The suppression is weaker at larger $r_L$, for which the resonant scale approaches the surviving large-scale fluctuations.

Supersonic turbulence mitigates the effect of damping. Decoupling increases $D_\parallel$ by a factor of a few to ten rather than by $\sim10^2$, and the strong reduction of $D_{\mu\mu}$ near $\mu_0\rightarrow1$ is absent. The three coupling states remain distinct, confirming that supersonic turbulence is still affected by damping, but their difference is substantially smaller because the shock-associated fluctuations preserve pitch-angle scattering. The intermediate curves lie between the coupled and decoupled limits. At $M_A\simeq1$, however, the transport remains highly anisotropic, with $D_\parallel/D_\perp\sim10^3$--$10^5$. Maintaining pitch-angle scattering is therefore necessary but not sufficient; nearly isotropic transport additionally requires decorrelation of the magnetic-field direction.

\begin{figure*}[t]
\centering
\includegraphics[width=0.99\linewidth]{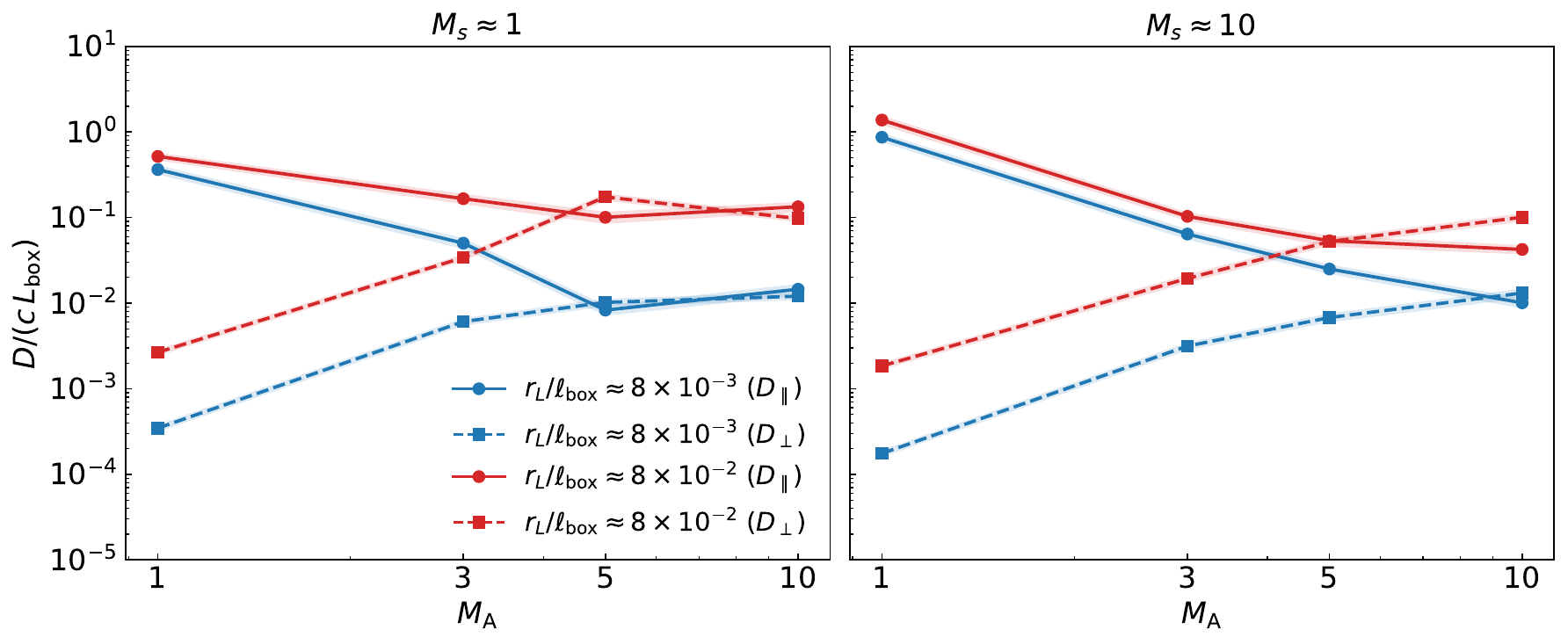}
\caption{\label{fig:ma_scan}
Dimensionless diffusion coefficients as functions of Alfv\'enic Mach number in solenoidally driven single-fluid turbulence. The left and right panels show $M_s\simeq1$ and $M_s\simeq10$, respectively. Solid and dashed curves denote $D_\parallel/(cL_{\rm box})$ and $D_\perp/(cL_{\rm box})$ for $r_L/L_{\rm box}\simeq8\times10^{-3}$ and $8\times10^{-2}$. Increasing $M_A$ reduces $D_\parallel$ and drives $D_\parallel/D_\perp$ toward unity.}
\end{figure*}

\subsection{Super-Alfv\'enic isotropization}
Fig.~\ref{fig:ma_scan} isolates the dependence on $M_A$ in the strongly coupled limit. For both $M_s\simeq1$ and $10$, increasing $M_A$ from order unity to $\simeq10$ reduces $D_\parallel$ by more than an order of magnitude while increasing $D_\perp$. The coefficients approach one another for $M_A\gtrsim5$, with the closest agreement in the supersonic $M_A\simeq10$ runs.

This trend follows from the scale hierarchy of super-Alfv\'enic turbulence. Above the Alfv\'en scale $\ell_A=L_{\rm inj}M_A^{-3},$ the turbulent kinetic energy exceeds the magnetic energy and the field is advected by nearly hydrodynamic motions. Below $\ell_A$, magnetic tension becomes dynamically important, and an anisotropic MHD cascade develops. For particles with a parallel mean free path $\lambda_\parallel\gtrsim\ell_A$, the local field direction decorrelates after a displacement of order $\ell_A$, yielding an approximately isotropic random walk with \citep{2008ApJ...673..942Y}
\begin{equation}
D_\parallel\simeq D_\perp\simeq\frac{1}{3}c\ell_A.
\label{eq:Deff}
\end{equation}
Increasing $M_A$ therefore shortens both the field-direction correlation length and the effective spatial step of the random walk. The transport becomes slower and less sensitive to the imposed mean field. Locally, particles can still propagate preferentially along an individual field segment, but the segment orientation changes repeatedly on scales above $\ell_A$. 

Eq.~\eqref{eq:Deff} predicts an energy-independent morphology in the sense of transport anisotropy once $\lambda_\parallel\gtrsim\ell_A$. If $\lambda_\parallel$ generally increases with particle energy, the gamma-ray halo's morphology produced by particles above the characteristic energy should consequently become approximately isotropic for a given source. Radiative cooling may still make the higher-energy emission more compact, but it does not by itself change the isotropic versus anisotropic character of the morphology.

In particular, lower-energy particles with $\lambda_\parallel\lesssim\ell_A$ remain more closely tied to the local magnetic-field direction and may produce a more anisotropic morphology. Future energy-resolved measurements, for instance by H.E.S.S. and LHAASO \citep{2022PhRvD.106l3017F, 2025arXiv251003183S}, could determine whether the halo axis ratio evolves with energy.

\begin{figure*}[t]
\centering
\includegraphics[width=0.99\linewidth]{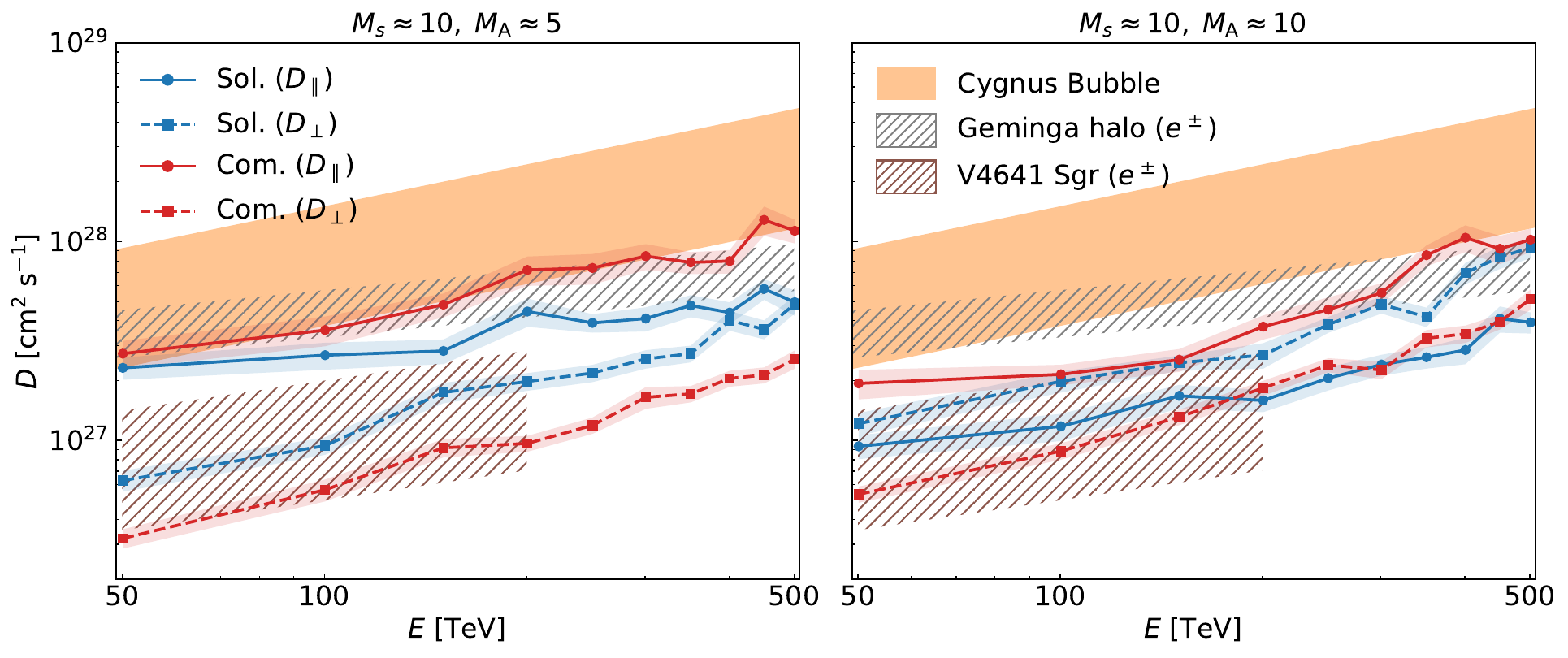}
\caption{\label{fig:physical_scaling}
Physical diffusion coefficients for $M_s\simeq10$ turbulence, adopting $L_{\rm box}=1\,\mathrm{pc}$. The left and right panels show $M_A\simeq5$ and $M_A\simeq10$, respectively. Solid and dashed curves denote $D_\parallel$ and $D_\perp$; blue and red denote solenoidal and compressive forcing. Shaded bands indicate representative diffusion ranges inferred for the Cygnus Bubble, the Geminga halo, and the V4641~Sgr microquasar.}
\end{figure*}

\section{Astrophysical Implications and Limitations}\label{sec:discussion}
\subsection{Slow and nearly isotropic diffusion}
 To illustrate the implications of our simulations, we adopt a fiducial scale of $L_{\rm box}=1\,\mathrm{pc}$, comparable to the magnetic-field coherence scale inferred in models of the Geminga halo \citep{2018MNRAS.479.4526L}. We further adopt $B_0=6.5\,\SI{}{\micro G}$ as a representative interstellar magnetic field strength \citep{2012ARA&A..50...29C}. These choices are not unique; the particle energies and diffusion coefficients scale with the adopted values of $B_0$ and $L_{\rm box}$. For these fiducial parameters, the simulated range $r_L/L_{\rm box}\simeq8\times10^{-3}$--$8\times10^{-2}$ corresponds to particle energies of approximately $50$--$500\,\mathrm{TeV}$, using
\begin{equation}
E\simeq0.93\,Z
\left(\frac{B_0}{\SI{}{\micro G}}\right)
\left(\frac{r_L}{\mathrm{pc}}\right)\mathrm{PeV},
\label{eq:energy_mapping}
\end{equation}
where $Z$ is the particle charge number; we take $|Z|=1$ throughout.

Fig.~\ref{fig:physical_scaling} compares the resulting diffusion coefficients with representative ranges inferred for the Cygnus Bubble, Geminga, and V4641~Sgr \citep{2017Sci...358..911A,2024SciBu..69..449L,2025ApJ...978L..20S}. These systems need not share the same emission mechanism or ambient plasma phase; they are included to illustrate the range of slow CR transport inferred in the vicinity of high-energy accelerators.

For $M_s\simeq10$ and $M_A\simeq5$--$10$, both solenoidal and compressive forcing yield diffusion coefficients of $10^{27}$--$10^{28}\,\mathrm{cm^2\,s^{-1}}$. The transport anisotropy, however, depends on both $M_A$ and the forcing geometry. This dependence is strongest at $M_A\simeq5$: compressive forcing retains $D_\parallel/D_\perp$ as large as $\sim10$, whereas solenoidal forcing produces a more disordered magnetic field and correspondingly more isotropic transport. At $M_A\simeq10$, the imposed mean field is dynamically weak, and the two forcing geometries converge toward slow, nearly isotropic diffusion. Strong compressions alone therefore do not ensure isotropic transport. At moderate $M_A$, shock-compressed magnetic structures can preserve a residual preferred direction. Slow and nearly isotropic diffusion emerges only when the scattering is efficient, and the mean magnetic field becomes subdominant dynamically.

\subsection{Ion-neutral damping effect}
The physical interpretation differs between source classes. For leptonic halos propagating through warm, highly ionized gas, the high-$M_A$ single-fluid calculations provide the more relevant limit for magnetic transport, although radiative losses must be included when predicting the electron distribution and gamma-ray surface-brightness profile. For hadronic emission surrounding candidate PeVatrons embedded in cold neutral or molecular gas, the two-fluid effects become essential: neutral--ion damping removes resonant-scale fluctuations in trans-sonic turbulence, but its impact is reduced when supersonic shocks preserve strong magnetic compressions. In both cases, large-scale isotropization is controlled primarily by the stochasticity of the magnetic field in super-Alfv\'enic turbulence, rather than by partial ionization itself.

The required conditions, $M_s\sim10$ and $M_A\sim5$--$10$, are more naturally associated with locally driven turbulence than with the quiescent interstellar cascade. Supernova shocks, stellar-cluster winds, and pulsar or microquasar outflows can inject motions that exceed both the sound and Alfv\'en speeds of the surrounding multiphase gas. The model therefore predicts that strongly suppressed diffusion should be localized around actively driven regions and should weaken as the turbulence decays. This picture is complementary to earlier superbubble models in which repeated shocks produce stochastic particle acceleration while spatial diffusion is prescribed phenomenologically \citep{1992MNRAS.255..269B}.

The model also predicts more elongated halos when a residual mean field remains dynamically important, as in the compressively driven $M_A\simeq5$ case. The influence of neutral--ion damping should vary with particle rigidity: as $r_L$ approaches the scale of the largest surviving magnetic compressions, the difference between coupled and decoupled transport diminishes, consistent with Fig.~\ref{fig:coupling}. Finally, because $\ell_A\propto M_A^{-3}$, even modest variations in the injection velocity can produce substantial source-to-source differences in both the diffusion coefficient and the halo morphology.

\subsection{Comparison with earlier work}
An alternative interpretation was proposed by \citet{2019PhRvL.123v1103L}, who showed that the approximately symmetric and compact appearance of the Geminga halo does not necessarily require intrinsically slow and isotropic diffusion. In their model, diffusion is strongly anisotropic in sub-Alfv\'enic turbulence, with the local mean magnetic field approximately aligned with the line of sight. The projected extent of the halo is then controlled mainly by the slower perpendicular diffusion, whereas rapid propagation along the mean field distributes particles over a larger line-of-sight depth. This geometry can therefore produce a compact, nearly circular projected halo without invoking the extreme turbulent conditions required for intrinsically isotropic transport.

Our results do not exclude this geometric interpretation. Indeed, it may be particularly relevant to evolved systems such as Geminga. As locally driven turbulence decays, the turbulent velocity and hence $M_A$ decrease, causing the mean magnetic field to become dynamically more important and the transport to become increasingly anisotropic. Our two-fluid results, however, introduce an additional environmental dependence that is absent from single-fluid models. In partially ionized gas, ion-neutral damping increases $D_\parallel$ while reducing $D_\perp$, thereby strengthening the contrast between transport along and across the mean field. It would consequently produce a smaller projected halo and a greater line-of-sight extent than predicted by an otherwise equivalent single-fluid turbulence model.

Distinguishing this geometry-driven scenario from intrinsically slow, nearly isotropic diffusion therefore requires joint modeling of the surface-brightness profile, angular extent, and total flux. In particular, one may test whether a two-fluid anisotropic model predicts a systematically smaller halo or a different line-of-sight-integrated brightness than its single-fluid counterpart for the same injection history and magnetic-field orientation.


\section{Conclusions}\label{sec:conclusion}
Ion-neutral damping generally accelerates CR transport along the magnetic field. In trans-sonic partially ionized turbulence, it removes the
gyroresonant fluctuations responsible for scattering loss-cone particles, thereby increasing $D_\parallel$ by approximately two orders of magnitude while further suppressing perpendicular transport. Supersonic turbulence remains affected by the damping, but shock-associated magnetic fluctuations continue to produce pitch-angle scattering and limit the increase in $D_\parallel$ to a factor of a few to ten.

Super-Alfv\'enic turbulence provides the complementary geometric effect. The magnetic-field direction decorrelates over the Alfv\'en scale
$\ell_A$, reducing transport along the global mean field and enhancing perpendicular transport until $D_\parallel$ and $D_\perp$ become comparable. For locally driven conditions with $M_s\simeq10$, $M_A\simeq5$--$10$, and the fiducial choice $L_{\rm box}=1\,\mathrm{pc}$, the resulting diffusion coefficients fall within the observationally inferred range of $10^{27}$--$10^{28}\,\mathrm{cm^2\,s^{-1}}$ around TeV--PeV accelerators. 

These results have direct implications for extended gamma-ray halos. Slow, nearly isotropic transport can retain energetic particles near their sources while producing approximately symmetric emission on the sky. The halo size and morphology should depend on both the local turbulent state and the ionization conditions: strongly super-Alfv\'enic regions should produce more nearly symmetric halos, whereas a dynamically important mean field should yield more elongated or viewing-angle-dependent morphologies. Ion-neutral dampingcan further increase this geometric sensitivity by enhancing parallel transport and suppressing perpendicular transport.

\begin{acknowledgments}
Y.H. thanks Chris McKee, James Stone, Alexandre Marcowith, and Mark Krumholz for helpful discussions. Y.H. acknowledges support from NASA through the NASA Hubble Fellowship grant No.~HST-HF2-51557.001 awarded by STScI, which is operated by AURA under NASA contract NAS5-26555. The Sherman Fairchild Postdoctoral Fellowship at the California Institute of Technology also supports Y.H.. This work used SDSC Expanse and NCSA Delta through ACCESS allocations PHY230032, PHY230033, PHY230091, PHY230105, PHY230178, and PHY240183, supported by NSF grants \#2138259, \#2138286, \#2138307, \#2137603, and \#2138296.
\end{acknowledgments}

\appendix
\section{Convergence study}

\begin{figure*}[t]
\centering
\includegraphics[width=0.7\linewidth]{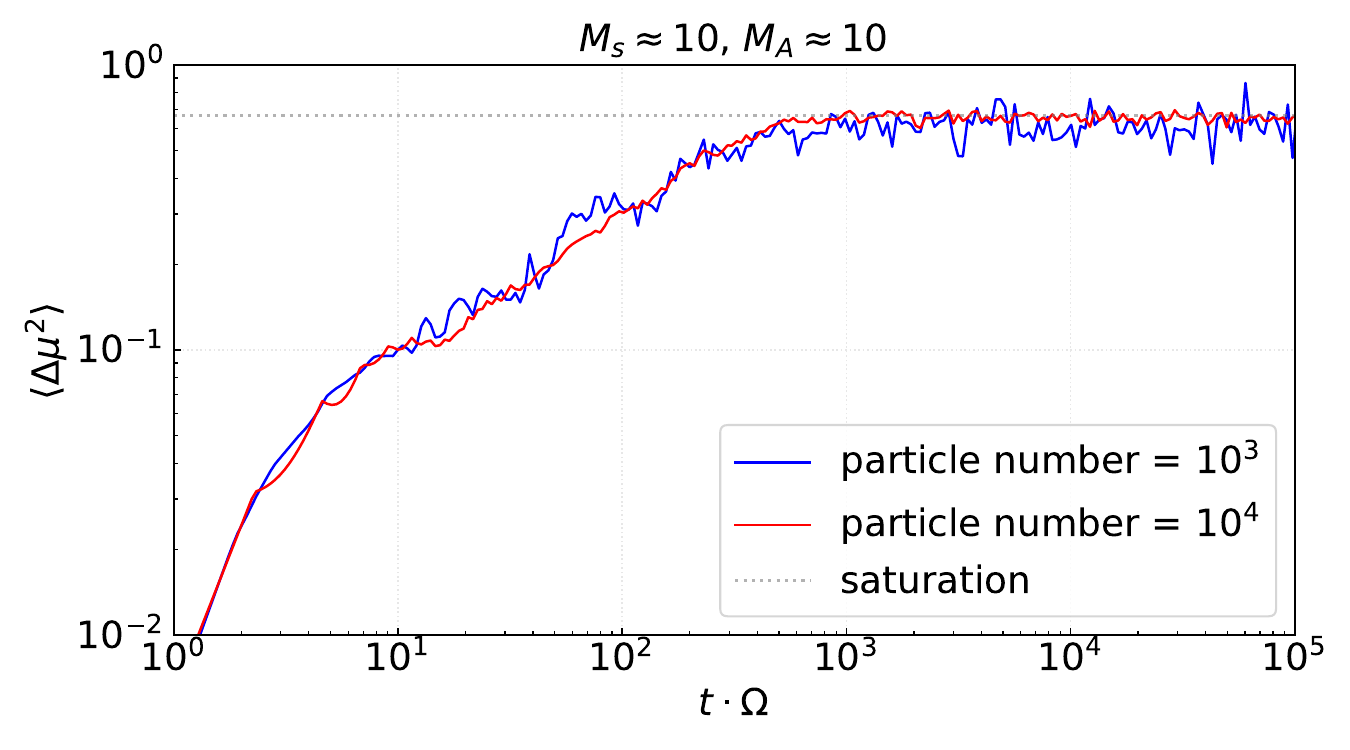}
\caption{\label{fig:convergence} Pitch-angle mean-square variations $\langle(\Delta\mu)^2\rangle$, as a function of the normalized time $t\Omega$ for the $M_{\rm s}\simeq10$ and $M_{\rm A}\simeq10$ simulation. The initial pitch angle distribution is isotropic with $r_L/L_{\rm box}\simeq8\times10^{-3}$. The blue and red curves show results obtained using $10^3$ and $10^4$ particles, respectively. The gray dotted line marks the saturation value. }
\end{figure*}

Because the pitch-angle cosine is bounded, $\mu\in[-1,1]$, the
pitch-angle mean-square variation cannot remain in a diffusive growth
regime indefinitely and instead approaches a finite plateau once the
particles lose memory of their initial pitch angles. We measure this
decorrelation using
\begin{equation}
    \left\langle (\Delta\mu)^2 \right\rangle
    =
    \left\langle
    \left[\mu(t)-\mu(0)\right]^2
    \right\rangle .
    \label{eq:mu_msd}
\end{equation}


We characterize the scattering strength by the saturation time $t_{\rm sat}$, defined as the characteristic time over which $\left\langle(\Delta\mu)^2\right\rangle$ approaches its late-time plateau. To exclude the initial non-diffusive evolution, we measure the growth from $t_0=10/\Omega$, after which $\left\langle(\Delta\mu)^2\right\rangle$ exhibits approximately linear growth in our simulations. We then define an effective pitch-angle diffusion coefficient as
\begin{equation}
    \frac{D_{\mu\mu}^{\rm eff}}{\Omega}
    \equiv
    \frac{
        \left\langle(\Delta\mu)^2\right\rangle_{t=t_{\rm sat}}
        -
        \left\langle(\Delta\mu)^2\right\rangle_{t=t_0}
    }{
        2(t_{\rm sat}-t_0)\Omega
    },
    \qquad
    t_0=\frac{10}{\Omega}.
    \label{eq:dmumu_sat}
\end{equation}

As shown in Fig.~\ref{fig:convergence}, we assess the numerical convergence of this estimate using ensembles of $10^3$ and $10^4$ particles. The two ensembles exhibit consistent growth and saturation of $\left\langle(\Delta\mu)^2\right\rangle$, demonstrating that the inferred decorrelation time and effective scattering rate are insensitive to the particle number over this range.

\bibliography{sample631}
\bibliographystyle{aasjournal}

\end{document}